\documentclass[manuscript,screen,review=false]{acmart}

\setcopyright{acmcopyright}
\copyrightyear{2023}
\acmYear{2023}
\acmDOI{XXXXXXX.XXXXXXX}

\acmJournal{JACM}
\acmVolume{37}
\acmNumber{4}
\acmArticle{111}
\acmMonth{8}

\acmPrice{15.00}
\acmISBN{978-1-4503-XXXX-X/18/06}

\usepackage[utf8]{inputenc}
\usepackage{graphicx}

\usepackage{bbm}
\usepackage[group-separator={,}]{siunitx}
\usepackage{multirow}
\usepackage[inline,shortlabels]{enumitem}
\usepackage[labelformat=simple]{subcaption}
\usepackage[noend]{algpseudocode}
\usepackage{algorithm}
\usepackage{algorithmicx}

\begin{document}

\title{A Survey on Incremental Update for Neural Recommender Systems}

\author{Peiyan Zhang}
\email{pzhangao@cse.ust.hk}
\affiliation{%
  \institution{The Hong Kong University of Science and Technology}
  \country{Hong Kong}}

\author{Sunghun Kim}
\email{hunkim@cse.ust.hk}
\affiliation{%
  \institution{The Hong Kong University of Science and Technology}
  \country{Hong Kong}
  }

\renewcommand{\shortauthors}{P. Zhang, et al.}

\begin{abstract}
    Recommender Systems (RS) aim to provide personalized suggestions of items for users against consumer over-choice. Although extensive research has been conducted to address different aspects and challenges of RS, there still exists a gap between academic research and industrial applications. Specifically, most of the existing models still work in an offline manner, in which the recommender is trained on a large static training set and evaluated on a very restrictive testing set in a one-time process. RS will stay unchanged until next batch retrain is performed. We frame such RS as Batch Update Recommender Systems (BURS). In reality, they have to face the challenges where RS are expected to be instantly updated with new data streaming in, and generate updated recommendations for current user activities based on the newly arrived data. We frame such RS as Incremental Update Recommender Systems (IURS).

In this article, we offer a systematic survey of incremental update for neural recommender systems. We begin the survey by introducing key concepts and formulating the task of IURS. We then illustrate the challenges in IURS compared with traditional BURS. Afterwards, we detail the introduction of existing literature and evluation issues. We conclude the survey by outlining some prominent open research issues in this area.
\end{abstract}




\maketitle

\section{Introduction}
\textbf{Recommender Systems (RS)}~\cite{ricci2011introduction} aim to provide personalized suggestions of items for users against consumer over-choice. Extensive research has been conducted to address the different aspects and challenges of RS. Generally speaking, various stuffs can be considered when generating recommendation results. For example, user interaction sequences, side information like item features and user features, contextual factors such as temporal timestamp~\cite{quadrana2018sequence} and spatial location~\cite{palmisano2008using}, etc. According to the different input data, the recommender models can be classified into content-based RS, collaborative filtering and hybrid RS~\cite{adomavicius2005toward}. 

With the development of deep learning, current recommender system research sees a proliferation of usages of deep learning techniques to develop more powerful neural recommender models. 
In academia, Hidasi et al.~\cite{Hidasi2016SessionbasedRW} first propose to leverage the recurrent neural networks (RNNs) to model users' preferences, and it was further enhanced with tailored ranking loss function for recommendation~\cite{Hidasi2018RecurrentNN}. Afterward, attention-based mechanisms are incorporated into the system and significantly boosted performance. NARM~\cite{Li2017NeuralAS} utilizes attention on RNN models to enhance the captured feature while STAMP~\cite{Liu2018STAMPSA} captures the long and short-term preference relying on a simple attentive model. Convolution Neural Networks (CNNs) are also leveraged in session-based recommendation. Tang et al.~\cite{tang2018personalized} try to embed item session as matrix and perform convolution on the matrix to get the representation. 
Yuan et al.~\cite{yuan2019simple} further improve the method with dilated convolutional layers, which helps in increasing the receptive fields. More recently, more developments focused on leveraging Graph Neural Networks (GNNs) ~\cite{Wu2019SessionbasedRW, Pan2020StarGN,Srivastava2015HighwayN,Wang2020PAGGANSR,Yu2020TAGNNTA,Wang2020GlobalCE,gupta2019niser, chen2020handling} to explore complex item transitions in session data. For example, Wu et al.~\cite{Wu2019SessionbasedRW} first propose to capture the complex transitions with graph structure. Afterward, Pan et al.~\cite{Pan2020StarGN} try to avoid overfitting through highway networks~\cite{Srivastava2015HighwayN}. Moreover, by empirically identifying the redundancy of the GNN-based models, Zhang et al.~\cite{zhang2023efficiently} remove the GNN propagation part and propose a more light-weight yet effective model with multi-level reasoning ability that captures the user preferences more accurately and efficiently.

In industry, neural recommender models are also widely deployed in most E-commerce and content systems such as Amazon, Netflix, Youtube and Google News~\cite{cheng2016wide,covington2016deep,davidson2010youtube,gomez2015netflix,okura2017embedding}. For example, Amazon and Alibaba~\cite{zhou2021contrastive} have deployed various recommender systems that serve billions of page views each day to identify and suggest products that customers are likely to purchase based on their preferences. YouTube's ~\cite{davidson2010youtube} home page recommendations accounted for 60\% of the clicks on videos, and Netflix's~\cite{gomez2015netflix} movie recommendations accounted for 80\%. With massive news articles emerging everyday, MSN News~\cite{wu2019neural} utilizes news recommender systems to capture user interests and generate personalized news recommendations, which helps users to find their interested news articles efficiently and alleviates information overload.


Despite the great success of neural RS have achieved nowadays, most of the existing models still work in an offline manner~\cite{mi2020ader}, in which the recommender is trained on a very large static training set and evaluated on a very restrictive testing set in a one-time process. RS will stay unchanged until next batch retrain is performed. We frame such RS as Batch Update Recommender Systems (BURS). In reality, they have to face the challenges where a recommender is expected to be instantly updated with new data streaming in, and generate satisfying recommendations for current user activities based on the newly arrived data. For example, if a user clicks an item, this should be reflected in the recommendation system immediately. However, recommender models ensuring a good recommendation quality tend to suffer from very high latency. Retraining the model regularly with the whole dataset is also infeasible. An enormous number of parameters creates a severe challenge to computing resources and data storage. Besides, if we only retrain the model on the newly collected data, a smaller amount of data from new users will make it difficult to generate satisfactory recommendations.

In particular, BURS has two main limitations. For one thing, the RS cannot take into account user interactions after a model update is performed until the next one is made, thus it cannot take continuous user activity into account. The same model is used to deliver recommendations, in spite of the fact that user preferences and item perceptions constantly change over time. Under this circumstance, recent observations are not taken into account instantly, resulting in poor recommendation quality~\cite{dias2008value}.
For another, periodic retraining becomes computationally expensive as the dataset grows over time. As a consequence, scalability issues are created. 


Research in the RS field has been focused on a portion of these limitations and addressed each of them using different approaches, i.e., time-aware RS~\cite{campos2014time} and continual learning RS~\cite{mi2020ader}. Nevertheless, these solutions are unable to overcome two of the limitations mentioned earlier, since they are not designed to handle new observations and model temporal dynamics simultaneously. Recent work~\cite{peng2021learning,zhang2020retrain} has considered to frame the problem of recommendation as a incremental update problem by designing efficient Incremental Update Recommender Systems (IURS), i.e., RS that can be instantly updated using only the newly arrived data to capture user interest drifts as well as item popularity changes.
IURS are capable of responding to changes in real time while operating independent of the number of observations, which addresses the previously mentioned limitations of BURS.

Overall, IURS are more realistic than traditional BURS in dynamic settings. Although existing work, i.e., the work reviewed in this survey, addressing various aspects of IURS, research in the field remains scattered and some important questions still remain unanswered. Our survey will review the existing works in IURS, discuss their differences from other RS families, and point out flaws in existing approaches.


The rest of this survey is organized as follows: we first introduce some background and formulate the task of IURS in Section~\ref{sec:problemDefinition}. Then, we illustrate the challenges in IURS compared with traditional batching RS in Section~\ref{sec:challenges}. Next, we detail the introduction to the existing literature and evaluation issues in Section~\ref{sec:relatedWork} and Section~\ref{sec:evaluation}, respectively. Finally, we discuss the prominent open research issues in Section~\ref{sec:futureDirections} and conclude this paper in Section~\ref{sec:conclusion}.

\section{Problem Definition} \label{sec:problemDefinition}
In this section, we first describe the classical settings of BURS to contrast the streaming settings of IURS. Afterward, we illustrate the streaming setting of IURS. Finally, to clearly define the problem considered, we give detailed algorithmic descriptions of BURS and IURS.

\subsection{Classical Settings of BURS}
Whereas RS aim to provide personalized suggestions of items for users against consumer over-choice, research in this topic requires to formulate this problem so that it can be quantified to evaluate. 
The first formulation is known as rating prediction, which focused on predicting the rating values that users tend to give on the unrated items. The second formulation is top-K recommendation, which aims to recommend the top-K item list that best matches the user's current preference. It is expected that the core of the RS should follow the same process for both cases~\cite{adomavicius2005toward}. Namely, the recommender models predict each item's utility, i.e., the rating score in the rating prediction or the relevance score in the top-K recommendation. Items for recommendation are based on the utility maximization.


Prior development of RS has mainly focused on the batch setting where the whole training dataset is available and static. Recommender models will be evaluated in a one-time process and remain static until retraining is required. There exists a huge gap between this practice and the real applications where recommender models are expected to be instantly updated with new data streaming in, and generate satisfying recommendations for current user activities based on the newly arrived data.  

When it comes to model user-item relationships in static domains where there are plenty of data available and the correlations are stable, there is no need to implement IURS that take new interactions into consideration as soon as they are generated. However, in real dynamic domains where user interests will drift and item popularites may change, it is important to instantly consider the new interactions that arrive at a high rate, which requires the IURS.


\subsection{Streaming Settings of IURS}

\textit{Streaming setting.} 
With data streaming in continuously and indefinitely, many new research challenges have been posed on analysis, computation and storage. Specifically, the following differences have been highlighted~\cite{article} in comparison with conventional BURS:
    \begin{itemize}
        \item New interactions arrive continuously.
        \item Data streams are possibly unbounded.
        \item As soon as an observation is processed from the stream, it is either discarded or reserved in a memory whose size is small compared to the size of the stream. Observations can only be retrieved when they have been reserved in the memory.
    \end{itemize}


\subsection{Algorithmic Descriptions}
In order to provide a clearer understanding of the problem, we describe here the algorithmic process followed by IURS in contrast to the traditional batch RS processes.

\begin{algorithm}[h]
  \caption{Algorithm of BURS}
  \label{al:1}
  \begin{algorithmic}[1]
    \Require
      $D^0$: Initial observed interaction set; $T$: model retraining time period;
    \State $D \leftarrow D^{0}$;
    \State $M \leftarrow f_{train}(D)$;
    \Function{Recommend}{}
    \For{each user $u$ at time $t$}
        \State $list=f_{predict}(M, u)$;
        \State Observe user $u$'s feedback $r_{ui}$ on item $i$;
        \State $D \leftarrow D\cup \{u, r_{ui}, t_{ui}, i\}$;
    \EndFor
    \EndFunction
    \Function{Retrain}{}
    \For{each time period $T$}
        \State $M \leftarrow f_{train}(D)$;
    \EndFor
    \EndFunction
    \State Recommend() \& Retrain(); 
  \end{algorithmic}
\end{algorithm}

\begin{algorithm}[h]
  \caption{Algorithm of IURS}
  \label{al:2}
  \begin{algorithmic}[1]
    \Require
      $D^0$: Initial observed interaction set; $T$: model retraining time period;
    \State $D \leftarrow D^{0}$;
    \State $M \leftarrow f_{train}(D)$;
    \State $D_{mem} \leftarrow f_{update}(D)$;
    \Function{Train}{$M, u, r_{ui}, t_{ui}, i, D_{mem}$}
    \State $M \leftarrow f_{incre-train}(M, u, r_{ui}, t_{ui}, i, D_{mem})$;
    \EndFunction
    \Function{Recommend}{}
    \For{each user $u$ at time $t$}
        \State $list=f_{predict}(M, u)$;
        \State Observe user $u$'s feedback $r_{ui}$ on item $i$;
        \State Train($M, u, r_{ui}, t_{ui}, i, D_{mem}$);
        \State $D_{mem} \leftarrow f_{update}(u, r_{ui}, t_{ui}, i)$;
    \EndFor
    \EndFunction
    \State Recommend(); 
  \end{algorithmic}
\end{algorithm}

\textit{Notations.} 
In order to illustrate the algorithms, we will first establish a set of notations. Let $U = \{u_1, u_2,...,u_{|U|}\}$ be the universal set of users that contains all users handled by the system and $I = \{i_1,i_2,...,i_{|I|}\}$ the universal set of items. We denote $D$ as the observed interaction set where each interaction consists of $(u, r_{ui}, t_{ui}, i)$. $r_{ui}$ is user $u$'s feedback on item $i$ while $t_{ui}$ denotes the timestamp when the interaction occurs. Note that there may be various user feedbacks on items such as view, click, rate, buy and so on, here we do not use extra notations to record the feedback type as they are not explored in this survey. 

Given the established notations, the procedures of algorithms of BURS and IURS are showed in Algorithm~\ref{al:1} and Algorithm~\ref{al:2}, respectively. For the BURS, recommender models generate predicted item list for user $u$ (Algorithm~\ref{al:1}, line 5). Once user $u$'s feedback on item $i$ is observed, this interaction is added to the whole dataset. However, before the next retraining procedure, the models remain unupdated. For a certain time period $T$, the models will be retrained using the whole dataset $D$ (Algorithm~\ref{al:1}, line 10), which is not scalable in real applicaitons. In contrast, for IURS, each time when recommender models generate recommendation lists for user $u$ (Algorithm~\ref{al:2}, line 8) and observe user $u$'s feedback (Algorithm~\ref{al:2}, line 9), models will be incrementally trained using data in the memory (Algorithm~\ref{al:2}, line 10). Afterward, the data memory will be updated based on the newly observed interactions (Algorithm~\ref{al:2}, line 11).


\section{Challenges} \label{sec:challenges}
The contrast between IURS and traditional BURS poses several challenges that an incremental update RS should handle properly. In this section, we will present these challenges in the following.
\subsection{System Dynamic Modeling}
Users' interests may drift significantly while item popularities will also change temporally over time on dynamic online RS. This poses an urgent need to model the system dynamic accurately, as many underlying objects are evolving at different rates and in different ways. BURS will stay unchanged and deliver the same recommendations between two retrain intervals, which is unable to consider recent observations instantly and results in poor recommendation quality. IURS are required to handle such dynamic settings and model the temporal correlations suitably.

\subsection{Incremental Learning}
Incrementally updating IURS is a long period of work. To handle IURS with long duration, previous methods try to learn representations of the whole historical data with a large memory\cite{mi2020memory} or preserve representative samples to retrain after some time\cite{qiu2020gag}. However, as data is streaming in infinitely, the size of the memory is limited that is unable to preserve all representative samples. Thus IURS should avoid the train-from-scratch paradigm but turn to the incremental learning. To achieve the incremental learning for IURS, we have to further tackle two challenging tasks. One is catastrophic forgetting that the model may forget previous knowledge it has learnt as the past knowledge learned will be ultimately overwritten by the new knowledge learned incrementally. The other is effectively transferring past knowledge that is useful for future recommendation.



\section{Related Work} \label{sec:relatedWork}
\subsection{System Dynamic Modeling}
In IURS, users' interests may drift significantly while item popularities will also change temporally over time. Therefore, it is important to take the temporal information into consideration. Several methods were proposed to integrate time into RS~\cite{campos2014time}. We can mainly distinguish between three categories of approaches merging time with RS. 

\subsubsection{Recency-based Time-aware Models}
Recency-based time-aware models assume that recent observations are more pertinent than the older ones since they are more representative of the actual reality: they should therefore be given more importance in the learning and recommendation processes.
Note that the timestamps here are just item orders or position numbers in sequences. Most sequential-based and session-based recommender models follow this idea. They tend to generate short-term preference based on the last few items and concatenate the representation with the long-term preference representation~\cite{he2021locker,Wu2019SessionbasedRW, Pan2020StarGN,Srivastava2015HighwayN,Wang2020PAGGANSR,Yu2020TAGNNTA,Wang2020GlobalCE,gupta2019niser}. Note that compared with the following time-aware models, recency-based time-aware models are quite simple yet effective in capturing user interest drifts. However, most of these models work in a static setting.

\subsubsection{Contextual Time-aware Models}
Contextual time-aware models integrate the time dimension by considering temporal information as a contextual dimension. 
Fan et al.~\cite{fan2021continuous} unify sequential patterns and temporal collaborative signals. To capture temporal effects, they concatenate user-item embedding with interaction time embedding. To generate the interaction time embedding, they model the time span (temporal correlation) as the dot product of corresponding encoded time embeddings. Therefore, the time encoding function can be generalized to any unseen timestamp such that any time span not found in training data can still be inferred by the encoded time embeddings. With the generated time embedding, they manage to 
exploring the temporal effects and uncovering sequential patterns among the data. The performance of the approach is evaluated on five datasets in the context of top-K recommendation. The proposed approach show better results in terms of Recall and MRR metrics compared to other time-aware approaches, including that of TiSASRec~\cite{li2020time}. 

To characterize the dynamics from both user side and item side, Chen et al.~\cite{chen2021learning} propose to build a global user-item interaction graph for each time slice and exploit time-sliced graph neural networks to learn user and item representations. However, on each time sliced, they treat these graphs in a static setting and aggregate information along such graphs. To show the temporal property, they adopt two GRUs to aggregate user/item representations along all time slices from user and item side, respectively. This system modeling method is still static in nature.

Adopting timestamps to model the influence of time-decaying is another common choice. Zhang et al.~\cite{zhang2019dynamic} build a hierarchical attention network, where they add time-decaying factor on the attention weights to reflect the dynamism. This idea is based on the observation that people tend to read similar news content within a short period of time. However, when time passed for a long period, this news exerted less impact on the user. Li et al.~\cite{li2017learning} encode time correlations into a product-product graph and calculate the user embedding based on the set of products purchased by the user before timestamp $t$ in the form of a time-decaying function. The time-decaying function ensures that the later the product is purchased, the larger the weight of the product is. Motivated by the memory decay in reinforcement learning, Zhou et al.~\cite{zhou2021temporal} use an exponential denominator to re-scale the edge weights. Therefore, the historical edge weight impact will become more smaller
with the increasing of time discrepancy, which models the edge decaying in dynamic systems.

\subsubsection{Continuous Time-aware Models}
Continuous time-aware models represent user interactions as a continuous function of time, and the model parameters are learned from the data. 

Bai et al.~\cite{bai2019ctrec} propose a demand-aware Hawkes process~\cite{laub2015hawkes} to capture the complex influence of previous items on the future demands. It is natural to build such a continuous-time model due to the fact that users demands are highly time sensitive: it may be generated by their recent purchases, as well as the purchases made some time ago, i.e. short-term and long-term demands. Afterwards, they model the short-term demands through a convolutional time-aware LSTM while capturing long-term demands by a self-attentive component.

Kumar et al.~\cite{kumar2019predicting} adopt coupled RNNs to jointly model evolution of users and items and predict the future embedding trajectory of the user. In addition, they propose the embedding projection operation that project the embedding of a user at a future time. They incorporate time into the projected embedding via Hadamard product. Along this way, the predicted user embedding would drift farther as more time elapses. When the next interaction is observed, the embedding is updated again. Note that this method is in fully continuous fashion that the model update is triggered by each interaction.

Choi et al.~\cite{choi2021lt} first introduce neural ordinary differential equations (NODEs) on continuous RS modeling. After redesigning linear GCNs~\cite{he2020lightgcn} on top of the NODEs regime, they manage to learn the optimal architecture rather than relying on manually designed ones as well as learning smooth ODE solutions that are considered suitable for collaborative filtering. Recently, Guo and Zhang et al.~\cite{guo2022evolutionary} propose Graph Nested GRU Ordinary Differential Equations (GNG-ODE) that further extends the idea
of NODEs to continuous-time temporal session graphs. GNG-ODE manages to preserve the continuous nature of dynamic user preferences, encoding both temporal and structural patterns of item transitions into continuous-time dynamic embeddings to capture the fine-grained user preference evolution.

\subsection{Incremental Learning}
To achieve the incremental learning for IURS, we have to further tackle two challenging tasks. One is catastrophic forgetting that the model may forget previous knowledge it has learnt as the past knowledge learned will be ultimately overwritten by the new knowledge learned incrementally. The other is effectively transferring past knowledge that is useful for future recommendation. 

In fact, the forgetting issue is not exclusive to RS incremental update, but generally exists when updating neural networks with data from a different distribution~\cite{kirkpatrick2017overcoming}. Continual learning is such a research topic that deals with the catastrophic forgetting problem to achieve lifelong learning, which has been successfully applied in Computer Vision~\cite{delange2021continual} and Natural Language Processing~\cite{biesialska2020continual}. However, when it comes to RS incremental update, directly adopting these methods may not lead to desirable outcomes. This is because the two problems have different ultimate goals: continual learning aims to perform well for the current task without sacrificing performance on previous tasks, while incremental update of RS only cares about the performance on future recommendations~\cite{zhang2020retrain}.

Recently there are some attempts combining continual learning with recommender systems with the hope of training a robust model which contains all the historical recommendation experience to facilitate new recommendation tasks. Specifically, two lines of works have been proposed as follows.

\subsubsection{Sample-based Approach}
The first line of works is sample-based approach, a method of storing a small set of representative sequences from previous data, namely exemplars. Formally speaking, it usually consists of two parts: Exemplar Set Update and Representation Update.

\textbf{Exemplar Set Update.} The element of the exemplars for each item is called the examples of this item, i.e., user interaction sequences that contains the interaction with this item. Once the examples of a new item $i$ appears, sample-based approach will balance the size of exemplars for each item. For example, the exemplar size of existing items will be reduced while a new exemplar set will be created for the new item $i$. Suppose the total exemplar size is $M$ due to the memory limitation, then each item could receive $m = \frac{M}{|I|}$ exemplar quota where $|I|$ is the size of the existing itemset. What's more, the exemplar quota of each item can also be proportional to the item's appearance frequency. Here we provide the intuition of how to choose the examples. The mean feature vector over the examples should be close to the mean feature vector of the corresponding item. Therefore, the most representative examples can be selected and the general property of this item can be preserved.

The Exemplar Set Update process is illustrated in Algorithm~\ref{al:3}. For each existing item, we first compute its mean feature vector based on the interaction sequences that contain this item (Algorithm~\ref{al:3}, line 2). Afterwards, we select $m$ examples based on the distance from the item mean feature vector (Algorithm~\ref{al:3}, line 3-4). Note that these $m$ examples are ordered based on this distance, which will be beneficial to reduce the exemplar size when new items emerge in the future. Eventually, we can obtain the updated exemplar set for each existing item (Algorithm~\ref{al:3}, line 6). Note that we will drop all interactions that are not in the exemplar.
\begin{algorithm}[h]
  \caption{ExemplarSet Update}
  \label{al:3}
  \begin{algorithmic}[1]
    \Require
      $S_i=\{s_1,s_2,...,s_n\}$: Interaction sequences that contain item $i$; $m$: the exemplar size;$\phi$:feature extractor that compute sequence representation based on the input interaction sequence;
    \Ensure
      $P_t$: Exemplar set for all items at time $t$;
    \For{$y$ in 1,...,$|I_t|$}
    \State $\mu \leftarrow \frac{1}{n}\Sigma_{s\in S_y}\phi(s)$;
    \For{$k$ in 1,...,$m$}
        \State $p_{k}\leftarrow argmin_{s\in S \&\& s\notin \{p_1,...,p_{k-1}\}}||\mu - \frac{1}{k}[\phi (s)+\Sigma^{k-1}_{j=1}\phi (p_j)]||$;
    \EndFor
    \State $P_{ti} \leftarrow (p_1,p2,...,p_m)$; 
    \EndFor
    \State $P_t \leftarrow (P_{t1},P_{t2},...,P_{t|I|})$;
    \\
    \Return $P_t$;
  \end{algorithmic}
\end{algorithm}

\textbf{Representation Update.} Once we finish constructing exemplar sets for the current dataset, sample-based approach would perform representation update operations. Algorithm~\ref{al:4} details the steps for updating the feature representation. Given the newly observed interactions at time $t$ and Exemplar set $P_{t-1}$ at time $t-1$, we first initialize the model with parameters at time $t-1$ (Algorithm~\ref{al:4}, line 1) and train it on the data $D_{t}\cup P_{t-1}$ (Algorithm~\ref{al:4}, line 2-3). Afterwards, we update the exemplar size at time $t$ with Algorithm~\ref{al:3}. Typically during the training process, knowledge distillation~\cite{hinton2015distilling} will be adopted to force the model not forgetting the previous knowledge.


\begin{algorithm}[h]
  \caption{Representation Update}
  \label{al:4}
  \begin{algorithmic}[1]
    \Require
      $D_t$: New observed interactions at time $t$; $P_{t-1}$: Exemplar set at time $t-1$; ${\theta}_{t-1}$: model parameters at time $t-1$;
    \Ensure
      ${\theta}_t$: Updated model parameters at time $t$; $P_{t}$: Updated exemplar set at time $t$;
    \State Initialize ${\theta}_t$ with ${\theta}_{t-1}$;
    \While {${\theta}_t$ not converged}
    \State Train ${\theta}_t$ on the data $D_{t}\cup P_{t-1}$;
    \EndWhile
    \State Compute $P_{t}$ using Algorithm~\ref{al:3};
    \\
    \Return ${\theta}_{t}, P_{t}$;
  \end{algorithmic}
\end{algorithm}

Following this idea, \cite{diaz2012real} maintains a reservoir to keep history, where the samples are randomly selected. \cite{chen2013terec,wang2018streaming,guo2019streaming,zhao2021stratified,qiu2020gag,mi2020ader,mi2020memory} also maintain reservoirs to keep the history. However, the samples are selected based on carefully designed heuristics. Sample-based approach can be seen as an intermediate between incremental update and training on full historical data, aiming to find a balance between short-term interest and long-term memory \cite{peng2021learning}.

Despite the great success sample-based approach has achieved on the public datasets, such approach is not suitable in practice. On the one hand, the reservoir/memory size is limited while the streaming data is infinite. In spite of careful selection of samples with a certain amount of informativeness, these selected samples still fall short of capturing a comprehensive picture of the data distribution. On the other hand, most of these methods directly replay samples stored in the reservoirs, which do not judge whether the data is useful for the current task.

\subsubsection{Model-based Approach}
Another line of works is called Model-based Approach which aims to prevent catastrophic forgetting by transferring past knowledge to current models. Through treating past models as the teacher model, Wang et al.~\cite{wang2020practical} regularize the learning process with a knowledge distillation loss to prevent catastrophic forgetting. Additionally, subsequent studies employ distillation techniques with specific attention to GNNs~\cite{xu2020graphsail} and session-based RSs~\cite{mi2020ader}. However, it is important to note that simply limiting the change in networks will not lead to transferring useful knowledge to current tasks. In other words, some of knowledge in past models might not be applicable for the current task. Zhang et al.~\cite{zhang2020retrain} introduce a method for explicitly optimizing the next period's data, which greatly alleviates the problem of \cite{wang2020practical,xu2020graphsail,mi2020ader}. To optimize for future use, it develops a transfer module to combine knowledge of different periods' models. Nevertheless, this method fails to take the long-term sequential patterns into consideration, as the model mainly considers the transfer between two consecutive periods. Peng et al.~\cite{peng2021learning} first train models for each period data. It designs a hyper model which takes parameters from the previous models as input and output a model with the fine-tuned parameters to serve for the current task. However, it requires the hyper model to fine-tune every parameter of the recommendation model, which is infeasible for large-scale recommendation models with massive amount of parameters.
\section{Evaluation} \label{sec:evaluation}
In this section, we review the evaluation methodologies and metrics for IURS. In contrast to conventional BURS, IURS takes the temporal correlations between interactions into consideration and model several evolving concepts such as user interest and item popularities, which results in a different framework for evaluation.

It is well known that there exists online and offline evaluations for RS~\cite{gunawardana2015evaluating}. However, in this section, we focus on reviewing the offline evaluations as online evaluation requires an online environment that serve to recommend for real users, which is difficult for us to achieve. Nevertheless, online evaluation remains an important area to explore as it can provide the most convincing evidence regarding the ultimate goal of RS. 


\subsection{Offline Evaluation of BURS}
Offline evaluation is one of the most widely accepted settings during the RS evaluation procedure due to its simplicity that does not interact with real users and high efficiency in replicability as well as comparisons between different approaches~\cite{gunawardana2015evaluating}. 

The basic offline evaluation framework lies on the widely used train-test as well as cross-validation process. Given the observed dataset, it will be split into training set which is used to train the model, and test set which is used to evaluate model's performance. There exist various strategies to achieve the above mentioned dataset separation such as random split, split by time and leave-one-out strategy~\cite{gunawardana2015evaluating}. 

Although the conventional batch evaluation, which involves hold-out methods, is widely used to evaluate RS, it faces several challenges in the data streaming environment~\cite{vinagre2015evaluation}. If the data is randomly split into training set and test set, the semantics in the time dimension will be ignored or even destroyed. What's more, shuffling data may lead to illogical operations such as using future observations to predict previous observations, which results in the severe data leak problem. Learning from the shuffled observations will pose a negative effect on capturing real temporal correlations between interactions. In addition, traditional batch evaluation assumes models remain static when making recommendations. However, IURS will keep updating as long as recommendations are generated.

\subsection{Offline Evaluation of IURS}
In contrast to traditional BURS evaluation, the data streaming setting poses new challenges for the appropriate evaluation for IURS. Various attempts have been introduced to solve this problem~\cite{matuszyk2014selective,vinagre2015evaluation,siddiqui2014xstreams,palovics2014exploiting} and most of them rely on the so-called prequential methodology~\cite{gama2013evaluating}.

\subsubsection{Evaluation Methodologies}
Different from the one-time process evaluation in traditional BURS, IURS are expected to perform the prequential evaluation. In prequential evaluation, each observation of user feedback is followed by an extra \textit{test-then-learn} procedure~\cite{vinagre2015evaluation}. Take the top-K recommendation as an example:
\begin{enumerate} [(i)]
    \item \textit{Predict:} Generate the predicted top-K item list for user $u$.
    \item \textit{Evaluate:} Evaluate the predicted item list after observe the true interaction of user $u$.
    \item \textit{Update:} Use the true interaction of user $u$ to update the recommender models.
\end{enumerate}


The prequential methodology enjoys several benefits. In addition to respecting the temporal relationships between observed interactions, it manage to track the RS performance as well as the system evolution continuously along time. Meanwhile, recommender models can also utilize the real-time statistics obtained from prequential evaluation to better perform user interest drift detection. However, it is not without its disadvantages as one offline evaluation method. The major limitation lies on the fact that the predicted item list is evaluated against only one item, which fails to explore other potential candidates that occur in the item list. Nevertheless, utilizing hybrid evaluation methods could alleviate this limitation to some extent. Siddiqui et al. \cite{siddiqui2014xstreams} propose to use a hybrid hold-out evaluation based on one part of data and prequential evaluation based on the rest data.


Before IURS are deployed online for service, they still require an initial offline warm-up training stage. Thus, the actual application environment is not completely a streaming setting. Therefore, offline evaluation should also model this kind of working process, which considers shifting from a batch mode into the streaming mode.
To deal with this challenge, the following evaluation protocol is proposed by Matuszyk et al.~\cite{matuszyk2014selective}, where the evaluation phase is divided into three stages:
\begin{enumerate} [(1)]
    \item The \textit{Offline Warm-up Training} stage. On this stage, about first 30\% of the data is split to train the model in a batch mode. The data used in this stage serves as the training set.
    \item The \textit{Batch-Test–Stream-Train} stage. On this stage, about next 20\% of the data is utilized to test the model trained in offline warm-up training stage. Note that after testing, the data is also used to incrementally update the models. The data used in this stage can also serve as the validation set.
    \item The \textit{Stream-Test-Stream-Train} stage. On this stage, the rest of the data is exploited to make the prequential evaluation~\cite{vinagre2015evaluation}.
\end{enumerate}

\begin{table}
\centering
  \caption{Public Datasets Used for Evaluating IURS}
  \label{tab:overall}
  \begin{tabular}{|c c c c c c c|}
    \hline
    \textbf{Dataset} & \textbf{User} & \textbf{Item} & \textbf{Events} \textbf{Domain} & \textbf{Reference} & \textbf{Used in Papers}  \\
    \hline
    Amazon Books & 2.5M & 929k & 12.8M & Book &~\cite{mcauley2013hidden}  &~\cite{frigo2017online,choi2021lt}\\
    Amazon Electronics & 884k & 96K & 1.3M & Electronic &~\cite{mcauley2013hidden} &~\cite{frigo2017online}\\
    Amazon Movies & 1.3M & 235k & 8.5M & Movie &~\cite{mcauley2013hidden}  &\cite{frigo2017online}\\
    Dating & 194k & 194k & 11.7M &Social Network &~\cite{brozovsky2007recommender}  &~\cite{subbian2016recommendations,wang2013online}\\
    Delicious & 1.8k & 69k & 105k &Point of interest &~\cite{brusilovsky2010workshop}  &~\cite{liu2014towards}\\
    Diginetica & 780k & 43k & 983k & E-commerce &~\cite{Wu2019SessionbasedRW} &~\cite{mi2020ader,zhou2021temporal}\\
    Douban & 36k & 726k & 5.3M &Social Network &~\cite{zhong2012comsoc}  &~\cite{liu2014towards}\\
    Flixster & 147k & 48k & 8.1M &Movie &~\cite{dong_2017}  &~\cite{matuszyk2017stream}\\
    Globo & 314k & 46k & 3M &News &~\cite{de2018news}  &~\cite{de2018news,gabriel2019contextual}\\
    Gowalla & 30k & 41k & 1M & Point of Interest &~\cite{cho2011friendship}  &~\cite{choi2021lt}\\
    Jingdong & 457k & 55k & 8.9M & E-commerce &~\cite{zhao2015connecting,zhao2014we}  &~\cite{bai2019ctrec}\\
    Last.fm-600k & 164 & 65k & 493k & Music &~\cite{aoscar2010music}  &~\cite{vinagre2015evaluation,vinagre2015collaborative,vinagre2018online}\\
    Last.fm-30M & 40k & 5.6M & 31M & Music &~\cite{turrin201530music}  &~\cite{frigo2017online,kumar2019predicting}\\
    Lazada & 10k & 44k & 6.7M & E-commerce &~\cite{peng2021learning}  &~\cite{peng2021learning}\\
    MovieLens-100k & 1k & 1.7k & 100k & Movie &~\cite{harper2015movielens}  &~\cite{al2018adaptive,chang2017streaming,matuszyk2017stream,wang2013online}\\
    MovieLens-1M & 6k & 4k & 1M & Movie &~\cite{harper2015movielens}  &~\cite{al2018dynamic,matuszyk2017stream,siddiqui2014xstreams,vinagre2015evaluation,vinagre2015collaborative,vinagre2018online,wang2018streaming}\\
    MovieLens-10M & 72k & 10k & 10M & Movie &~\cite{harper2015movielens}  &~\cite{al2018dynamic,chang2017streaming,frigo2017online,song2019coupled}\\
    MovieLens-20M & 138k & 27k & 20M & Movie &~\cite{harper2015movielens}  &~\cite{wang2018neural,peng2021learning}\\
    Netflix & 480k & 17k & 100M & Movie &~\cite{chang2017streaming}  &~\cite{chang2017streaming,matuszyk2017stream,song2019coupled,wang2018neural,wang2018streaming}\\
    Plista & 220k & 835 & 1.1M &News &~\cite{kille2013plista}  &~\cite{al2018adaptive,jugovac2018streamingrec,lommatzsch2015real}\\
    Reddit post & 10k & 1k & 672k &User posts &~\cite{kumar2019predicting}  &~\cite{kumar2019predicting}\\
    Sobazaar & 17k & 25k & 843k & E-commerce &~\cite{peng2021learning}  &~\cite{peng2021learning}\\
    Ta-Feng & 26k & 24k & 818k & E-commerce &~\cite{wang2015learning}  &~\cite{bai2019ctrec}\\
    Taobao & 26k & 24k & 818k & E-commerce &~\cite{zhu2018learning,zhu2019joint,zhuo2020learning}  &~\cite{bai2019ctrec}\\
    Tmall & 50k & 44k & 3M & E-commerce &~\cite{peng2021learning}  &~\cite{peng2021learning}\\
    Twitter & 413k & 37k & 35M &User posts &~\cite{yang2011patterns}  &~\cite{diaz2012happening,diaz2012real}\\
    Wikipedia edits & 8k & 1k & 157k &User posts &~\cite{kumar2019predicting}  &~\cite{kumar2019predicting}\\
    Yahoo! Music-6k & 6k & 127k & 476k &Music &~\cite{vinagre2018online}  &~\cite{vinagre2018online}\\
    Yelp2018 & 32k & 38k & 1.6M & Point of interest &~\cite{he2020lightgcn}  &~\cite{choi2021lt}\\
    \hline
  \end{tabular}
\end{table}

\subsubsection{Evaluation Metrics}
Prequential evaluation metrics are evaluated based on every observed user feedback. We can obtain the performance evolution along the time. The overall performance will be reported through averaging over all observed feedbacks. We present some metrics that are widely used in prequential evaluation as follows~\cite{frigo2017online}. Note that $rank_{u}(i)$ means the ranking position of item $i$ in the generated top-K item list for user $u$.
    \begin{itemize}
        \item The \textit{precision} score measures the portion of corrected recommended items during the test process:
        \begin{equation}
                Precision@K(u,i)=
                \begin{cases}
                0& \text{if $rank_{u}(i) > K$}\\
                \frac{1}{K}& \text{otherwise.}
                \end{cases}
        \end{equation}
        \item The \textit{recall} score represents whether the top-k recommendation list contains the target item:
        \begin{equation}
                Recall@K(u,i)=
                \begin{cases}
                0& \text{if $rank_{u}(i) > K$}\\
                1& \text{otherwise.}
                \end{cases}
        \end{equation}
        \item The \textit{Discounted Cumulative Gain (DCG)} score measures the quality of ranking:
        \begin{equation}
                DCG@K(u,i)=
                \begin{cases}
                0& \text{if $rank_{u}(i) > K$}\\
                \frac{1}{\log_{2}(rank_{u}(i)+1)}& \text{otherwise.}
                \end{cases}
        \end{equation}
        \item The \textit{Mean Reciprocal Rank (MRR)} score measures the position of the target item in the top-k recommendation list. Large MRR@K score means the target item stays in the top position of the top-k recommendation list. If the target item is not in the recommendation list, MRR@K score will be set to zero:
        \begin{equation}
                MRR@K(u,i)=
                \begin{cases}
                0& \text{if $rank_{u}(i) > K$}\\
                \frac{1}{rank_{u}(i)}& \text{otherwise.}
                \end{cases}
        \end{equation}
    \end{itemize}


\subsubsection{Datasets}
Current research in RS also sees a proliferation of public datasets that covers various domains in academic environments. However, not all of them can be directly adopted in the IURS setting. In fact, the datasets used for IURS generally need to meet the following two conditions. Firstly, there should be specific timestamps for observed interactions, otherwise the problem will be reduced to only consider the position orders, which ignores the time span and thus fails to express temporal effects. Secondly, the timestamp labeled on each interaction should indicate the actual time when the interaction happened, i.e., the time when user $u$ watches the movie $i$, not the time when user $u$ give likes or dislikes feedback on movie $i$. Despite the fact that some of such datasets have been utilized in IURS, some researchers~\cite{al2021survey} doubt that they are actually not suitable for the particular problem of IURS.

Table~\ref{tab:overall} provides a list of publicly available benchmark datasets as well as their statistics, which are commonly utilized in the literature of IURS. 
Note that the IURS column indicates whether the dataset is actually suitable in IURS settings as mentioned above. 
For example, datasets like \textit{MovieLens} and \textit{Netflix} are obviously not feasible in IURS as the timestamp labeled on the item is the time when user generate the feedback, not the actual time when user watch the movie. Under this circumstance, the temporal order among the interactions can not be guaranteed. On the contrary, a music-artist dataset that is used for music interest recommendation: \textit{Last.fm} is more appropriate due to its real-time feedback property. 

\section{Future Research Directions} \label{sec:futureDirections}
Incremental update of recommender systems is one of the key tasks in several online services, which aims to instantly update models with new data streaming in, and generate satisfying recommendations for current user activities based on the newly arrived data. While this survey reviews previous IURS research, several prominent open research issues still remain, some of which are noted in the following.

\subsection{Dynamic System Modeling} 
The user-item interaction networks of large E-commerce platforms are dynamic in nature, in which coupled items exhibit complex behavior as time progresses. Such systems tend to share the following features:
\begin{enumerate}
    \item The node itself has self-evolution. i.e. iPhone 8 in 2017 stands for up-to-date device while in 2019 stands for budget-priced device.
    \item The node will be influenced by its neighbors in the graph.
    \item The edge will be influenced by the features of source and target nodes.
    \item The edge itself also has self-evolution. i.e. The interactions between two items will change naturally due to some seasonal factors (e.g. holidays, festivals), which is not necessarily related to the change of item features.
\end{enumerate}

Learning good representations that reflect the node role in networks is a great tool for analyzing and mining user interests drift on dynamic systems. With the remarkable success of recent advances of deep learning on graphs, there have been renewed interests in network representation learning parameterized by deep neural networks. \cite{Perozzi:2014:DOL:2623330.2623732,zhou2018dynamic,ijcai2019-640}. However, most of the network representations learning methods are designed in static setting and can not be adapted to fast changing dynamic networks. In order to adapt to the dynamic network environment, many dynamic network representation learning methods have been developed\cite{DBLP:conf/www/WuZMGSH20,DBLP:conf/wsdm/SankarWGZY20,DBLP:conf/kdd/YuCAZCW18,trivedi2018dyrep}. Most of these methods regard the dynamic network as a collection of snapshots at different time points. Each snapshot is composed of nodes and edges during a period of time. In representation learning, these methods generate embeddings for snapshot of time $t$ according to the information of the snapshot at time $t$ and previous snapshots. This approach can only roughly estimate the changes of node embeddings in latent space between different snapshots while unavoidably ignore the continuity of changes of node states. Fan et al.~\cite{fan2021continuous} concatenate user/item embedding with the generated continuous-time embedding to unify sequential patterns and temporal collaborative signals. Choi et al.~\cite{choi2021lt} first apply neural ordinary differential equations (NODEs) on dynamic RS modeling. Guo and Zhang et al.~\cite{guo2022evolutionary} further extend the idea of NODEs to continuous-time temporal session graphs. Their performance improvement over baselines is significant, which sheds light on modeling the system dynamic in a fully continuous fashion.

\subsection{Incremental Learning}
Researchers have proposed two mainstream continual learning approaches to overcome the specific forgetting issues in RS incremental update, namely, sample-based approach and model-based approach. In spite of both approaches' effectiveness, there are some major limitations. For sample-based approach, although it aims to select and preserve the most representative samples, these selected samples still fall short of capturing a comprehensive picture of the data distribution. Model-based approaches, which accounts for knowledge transfer from previous and current models are proposed to solve the incremental update problem. However, one major limitation of the current model-based approaches is that, most of them rarely consider whether the past knowledge is applicable or not for the current tasks. Therefore, instead of directly retrain the model with all past representative knowledge, future effort should identify whether the past knowledge is applicable to the current task. ~\cite{french1999catastrophic} suggests that reducing the overlapped representations would reduce the interference that leads to catastrophic forgetting. Recent research in neural biology~\cite{yu2014sparse} also support this claim: An activated neuron would suppress the activity of its weaker neighbors through "lateral inhibition". With this progress, the brain has a powerful de-correlated and compact representation with minimal interference from different input patterns~\cite{aljundi2019selfless}. Hence, learning de-correlated representations might be helpful for combating catastrophic forgetting. In addition, with the idea of aligning gradients across tasks that has been proposed before~\cite{lopez2017gradient}, a non-trivial connection to a meta-learner can be established that allows them to achieve the same goal to further de-correlate the unrelated learned representations while encourage the knowledge transfer among related learned representations. In the future, such approaches could be extended to IURS.

\subsection{Retrieval Module} 
Many existing methods predict user-item interactions by scoring all items for each user~\cite{beutel2018latent,dai2016deep,zhang2019deep,zhou2018dynamic}, which has a linear time complexity. This prevents such methods from scaling to datasets with millions of interactions.
Kumar et al.~\cite{kumar2019predicting} generate the embedding of predicted item instead of an interaction distribution. They recommend the item that is closest to the predicted item in the embedding space. Combined with local sensitive hashing method, they manage to reduce the retrieve time from linear time complexity to a constant time. Zhou et al.~\cite{zhou2021contrastive} propose to store item representations that have been computed in previous batches into a queue. They then adopt the queue to be an effective sampler for negative samples. Due to the computation reuse, such queue-like design achieves significant efficiency in deep candidate generation. This method has been deployed on line to serve billions of page views each day in Taobao. However, how to efficiently retrieve recommendation lists still  remains an open issue.


\subsection{Evaluation of IURS} 
Previous work related to IURS only considers the evaluation of recommendation accuracy with metrics like Precision or MRR scores. However, other issues such as item diversity, recommendation fairness are also critical in the online platform. It could also be interesting to explore the bias problems in the RS evaluation procedure~\cite{joachimsrecommendations,yang2018unbiased}. For example, recommender models induce a bias towards popular items~\cite{morik2020controlling}. Popular items are more likely to collect additional feedbacks, which in turn may influence future recommendation and promote misleading rich-get-richer dynamics. Under this circumstance, models are biased towards recommending popular items, and fail to recommend relevant long-tail items (less popular or less frequent items). How to handle the so-called popularity bias problem is quite practical in the RS incremental update process.



\subsection{Connection with Other Areas in RS}
Another research direction could be introduce connections with other areas in RS. For example, it is worth trying to consider incorporating rich side information like hybrid RS~\cite{al2018adaptive}, cross-domain RS~\cite{cantador2015cross} and context-aware RS~\cite{adomavicius2011context,du2019sequential}. In addition, one can keep track with recent continual learning research to focus on solving train-from-scratch problem. For example,~\cite{mi2020ader} first introduce continual learning into session-based RS, which is an important step towards bridging the gap between batching RS and IURS. Future advances could also consider the design of such methods.

\section{Conclusion} \label{sec:conclusion}
In this article, we offer a systematic survey of incremental update for neural recommender systems. We introduce key concepts and formulating the task of Incremental Update Recommender Systems (IURS). We then illustrate the challenges in IURS compared with traditional Batch Update Recommender Systems (BURS). Additionally, we organize and cluster existing literature based on the specific challenges they deal with. Afterwards, we detail the illustration of evaluation issues in IURS. We also highlight a bunch of open problems and promising future extensions. Incremental update of neural recommender systems shares significant practical values in real industrial applications. We hope this survey can provide readers with a comprehensive understanding of the key aspects of this field, clarify the most notable advances, and inspire more research in this practical research field.

\bibliographystyle{ACM-Reference-Format}
\bibliography{sample-base}

\end{document}